\documentclass[twocolumn,showpacs,preprintnumbers,amsmath,amssymb,epsfig,widetext]{revtex4}

\usepackage{graphicx}% Include figure files
\usepackage{dcolumn}% Align table columns on decimal point
\usepackage{bm}% bold math
\usepackage{epsfig}

\newcommand{\la}{\lesssim}
\newcommand{\ga}{\gtrsim}

\input epsf

\def\be{\begin{equation}}
\def\ee{\end{equation}}
\def\ba{\begin{eqnarray}}
\def\ea{\end{eqnarray}}

\newcommand{\epspar}{\epsilon}
\newcommand{\thepar}{\theta}
\newcommand{\Bpar}{B}
\newcommand{\Epar}{E}

\begin{document}

\preprint{}

\title{The Large Scale Structure of $f(R)$ Gravity}

\author{Yong-Seon Song,$^{1}$ Wayne Hu,$^{1,2}$ and Ignacy Sawicki$^{1,3}$}
\email{ysong@cfcp.uchicago.edu}
\affiliation{{}$^1$ Kavli Institute for Cosmological Physics, Enrico Fermi
Institute,  University of Chicago, Chicago IL 60637 \\
{}$^2$ Department of Astronomy \& Astrophysics,  University of Chicago, Chicago IL 60637\\
{}$^3$ Department of Physics,  University of Chicago, Chicago IL 60637
}

\date{\today}

\begin{abstract}
We study the evolution of linear cosmological
 perturbations in $f(R)$ models of accelerated expansion
 in the physical frame where the gravitational dynamics
are fourth order and the matter is minimally coupled.  These models
predict a rich and testable set of linear phenomena.
For each expansion history, fixed empirically by cosmological distance measures,
there exists two
branches of $f(R)$ solutions that are parameterized by $B \propto d^{2}f/dR^{2}$.
For $B<0$, which include most of the models
previously considered, there is a short-timescale instability at high curvature
that spoils agreement with high redshift
cosmological observables.
For the stable $B>0$ branch, $f(R)$ models can reduce the large-angle
CMB anisotropy, alter the shape of the linear matter power spectrum,
and qualitatively change the correlations between the CMB and galaxy surveys.
All of these phenomena are accessible with current and future data and
provide stringent tests of general relativity on cosmological scales.
\end{abstract}

%\pacs{draft}

%\keywords{CMB-inflation}

\maketitle

\section{Introduction}

Cosmic acceleration can be explained either by missing energy with an exotic
equation of state, dubbed dark energy, or by a modification of gravity on large
scales.  Indeed the cosmological constant can be considered either as a constant
added to the Einstein-Hilbert action or as vacuum energy.  Non-trivial modifications
where the addition is a non-linear function $f(R)$ of the Ricci scalar  that becomes
important only at the cosmologically low values of $R$ have also been shown to
cause acceleration  \cite{Caretal03,CapCarTro03,NojOdi03}.  They are furthermore
free of ghosts and other types of instabilities for a wide range of interesting
cases  \cite{NojOdi03,Dic04,DeFHinTro06}.

Solar-system tests of gravity provide what is perhaps the leading challenge to
$f(R)$ models as a complete theory of gravity \cite{Chi03}.
The equivalence of $f(R)$ models to scalar-tensor theories lead to conflicts with
parameterized post-Newtonian constraints
at a background cosmological density of matter.
It is however still controversial whether the whole
class of $f(R)$ modifications can be ruled out by this equivalence.   Matter in
the solar system becomes non-minimally
coupled in the transformed frame leading to non-trivial modifications of the
scalar field potential.  In the original Jordan---or physical---frame, it has been shown that the
Schwarzschild  metric solves the modified Einstein equations of a wide range of
 $f(R)$ models \cite{BreNojOdiVan04,Multamaki:2006zb} but this solution is not
 necessarily relevant for the solar system \cite{EriSmiKam06}.  
 Recent work has also raised the question as to whether 
 solar-system gravity problems may become tractable if $f(R)$
 is viewed as simply a first-order
 correction term to the high $R$ limit of general relativity
 \cite{Cem05,Sot05,ShaCaiWanSu06,Far06}.

Regardless of the outcome of small-scale tests of gravity in $f(R)$ models, it
is worthwhile to examine the cosmological consequences of treating $f(R)$ as
 an effective theory valid for a cosmologically appropriate range of curvatures.
 At the very least by making concrete predictions of cosmological phenomena in
 these models, one gains insight on how cosmology can test gravity at the largest scales.

In this Paper, we develop linear perturbation theory for predicting cosmological observables
such as the Cosmic Microwave Background (CMB) and the large-scale structure of the universe
exhibited in galaxy surveys.
We work in the physical frame where the matter is minimally coupled and obeys simple
conservation laws.

We begin in \S \ref{sec:expansion} by reviewing the properties
of $f(R)$ models and their relationship to the expansion history of the universe.
In \S \ref{sec:linear}, we derive the fourth order perturbation equations and, using
general properties demanded by energy-momentum conservation \cite{Bar80,HuEis99,Ber06}, recast them into
a tractable second order form.  In \S \ref{sec:stability} we identify a short time scale
instability that renders a wide class of $f(R)$ models  not viable cosmologically.
We present solutions on the stable branch in \S \ref{sec:evolution} and explore
their impact on cosmological power spectra in \S \ref{sec:observables}.   We
discuss these results in \S \ref{sec:discussion}.

\section{Expansion History}
\label{sec:expansion}

We consider  a modification to the Einstein-Hilbert
action of the form \cite{Sta80}
\ba
S=\int d^4x \sqrt{-g}\left[ {R+f(R) \over 2 \mu^2}+{\cal L}_{\rm m} \right]\,,
\ea
where $R$ is the Ricci scalar,  which we will sometimes refer to as
the curvature, $\mu^{2}\equiv 8\pi G$, and ${\cal L}_m$ is the matter Lagrangian.
Variation % of the given Lagrangian
with respect to the metric yields the modified Einstein equations
\ba\label{eq:metricvar}
%&&
G_{\alpha\beta} +
f_{R} R_{\alpha\beta}-({f\over2} -\Box f_{R}) g_{\alpha\beta}
%\nonumber\\&&\quad
- \nabla_{\alpha}\nabla_{\beta}f_{R}
% \nonumber\\ &&
= \mu^{2} T_{\alpha\beta}\,,
\ea
where
$f_{R}\equiv df/dR$ and likewise $f_{RR}\equiv d^{2}f/dR^{2}$ below.
We  define the
metric to include scalar linear perturbations around
a flat FRW background in the Newtonian or longitudinal gauge
\begin{equation}
ds^{2} = -(1+2\Psi)dt^{2} + a^{2}(1+2\Phi)dx^{2} \,.
\end{equation}
Given that the expansion history and dynamics of linear perturbations
are well-tested in the high curvature, high redshift limit by the CMB,
we restrict our considerations to models that satisfy
 \cite{Caretal03,CapCarTro03,NojOdi03}
\begin{equation}
\lim_{R\rightarrow \infty} f(R)/R \rightarrow 0 \,.
\end{equation}
With this restriction, the main modifications for viable models with
stable high curvature limits
 arise well after the radiation
becomes a negligible contributor to the stress energy tensor.
We can then take it to have the matter-dominated form
\begin{eqnarray}
T^0_{\hphantom{0}0} &=& -\rho(1+\delta) \,, \nonumber\\
T^0_{\hphantom{0}i} &=& \rho\partial_i q\,, \nonumber\\
T^i_{\hphantom{i}j}  &=& 0 \,.
\label{eq:Tmunu}
\end{eqnarray}

The modified Einstein equations with the FRW background
metric and $\delta=q=0$ yields the modified Friedmann equation
\begin{equation}
H^2 - f_R (H H' + H^2) + {1\over 6} f + H^2 f_{RR} R' = {\mu^2 \rho \over 3} \,.
\end{equation}
Here and throughout, primes denote derivatives with respect to $\ln a$.

There is sufficient freedom in the function $f(R)$ to reproduce any
desired expansion history $H$.   Hence the expansion history alone
cannot be used as a test of general relativity though it can rule out
specific forms of $f(R)$.  The dynamics of linear perturbations on the other hand
do test general relativity as we shall see.

We therefore seek to determine a family of $f(R)$ functions that is
consistent with a given expansion history \cite{Multamaki:2005zs,CapNojOdiTro06,Nojiri:2006gh,NojOdi06,delaCruz-Dombriz:2006fj}.
 Without loss of generality, 
 we can parameterize the expansion history in terms of an equivalent dark energy
model
\begin{equation}
H^2 = {\mu^2 \over 3} (\rho+ \rho_{\rm DE})\,.
\end{equation}
This yields a second order differential equation for $f(R)$
\begin{equation}
 - f_R (H H' + H^2) + {1\over 6} f + H^2 f_{RR} R' = - {\mu^2 \rho_{\rm DE} \over 3}\,,
\end{equation}
where $H^2$ and $R$ are fixed functions of $\ln a$
given the matching to the dark energy model.

For convenience, let us define the dimensionless quantities
\begin{equation}
\Epar = {H^2 \over H_0^2}\,, \quad
{R \over H_{0}^{2}} = 3 (4\Epar + \Epar' ) \,, \quad
y = {f \over H_0^2}\,.
\end{equation}
Here $H_0 \equiv H(\ln a=0)= h/2997.9$ Mpc$^{-1}$ is the Hubble constant.
The modified Friedmann equation can be recast into an inhomogeneous
differential equation for $y(\ln a)$
\begin{eqnarray}
L[y] & = & -{\mu^2} {\rho_{\rm DE} \over H_0^2}
\left(  {4 \Epar'  +  \Epar'' \over \Epar}  \right) \,,
\end{eqnarray}
where the differential operator on the lhs is given by
\begin{eqnarray}
L[y] &\equiv& y'' -\left( 1 + {1 \over 2} {\Epar' \over \Epar} + {4\Epar'' + \Epar''' \over 4 \Epar' +  \Epar''} \right)y'
\nonumber\\
&& +
{1 \over 2} \left(  {4 \Epar'  +  \Epar'' \over \Epar}  \right) y\,.
\label{eq:backmatch}
\end{eqnarray}
For
illustrative purposes, we take an expansion history that matches a dark energy
model with a constant equation of state $w$,
\begin{equation}
\Epar = (1-\Omega_{\rm DE}) a^{-3} + \Omega_{\rm DE} a^{-3(1+w)}\,,
\end{equation}
and
\begin{equation}
{\mu^2 \over 3} {\rho_{\rm DE} \over H_0^2} = \Omega_{\rm DE} a^{-3(1+w)}\,.
\end{equation}
Since Eq.~(\ref{eq:backmatch}) is a second-order differential equation, the
expansion history does not uniquely specify $f(R)$ but instead allows
a family of solutions that are distinguished by initial conditions.  This additional freedom
reflects the fourth-order nature of $f(R)$ gravity.

To set the initial conditions, take $y_{\pm}$ to be the
two solutions of the homogeneous equation $L[y]=0$.
At high curvature, these solutions are power laws $y_\pm \propto a^{p_\pm}$ with
\begin{equation}
p_\pm = {-7 \pm \sqrt{73} \over 4} \,.
\end{equation}
Since $p_- \approx -3.9$, stimulation of this decaying
mode violates the condition that $f_R/R \rightarrow 0$
at high $R$.   We therefore set its amplitude to zero in our solutions (c.f.
\cite{CapNojOdiTro06,AmePolTsu06a,AmePolTsu06b}).
The particular solution in the high curvature limit becomes
\begin{eqnarray}
 y_{\rm part} &=& {6 \Omega_{\rm DE} \over 6w^{2 }+ 5 w -2} a^{-3(1+w)} \,.
\end{eqnarray}
Therefore when numerically integrating Eq.~(\ref{eq:backmatch}) we take
\begin{eqnarray}
y(\ln a_i) &=& A y_+(\ln a_i) + y_{\rm part}(\ln a_i)  \,, \\
y'(\ln a_i) &=& p_+ A y_+(\ln a_i) - 3(1+w)y_{\rm part}(\ln a_i) \nonumber \,,
\label{eq:backgroundfR}
\end{eqnarray}
at some initial epoch $a_i \sim 10^{-2}$.

Since the modifications to gravity appear at low redshifts, it is more convenient
to parameterize the individual solutions in the family
by the final conditions rather than the growing
mode amplitude of the initial
conditions.  Given that a constant $f(R)$ is simply a cosmological constant and
a linear one represents  a rescaling of $G$ or $\mu$, it is $f_{RR}$, the second derivative,
that controls phenomena that are unique to the modification.  In particular, we
shall see that a specific dimensionless quantity
\begin{eqnarray}
\Bpar &=& {f_{RR} \over 1+f_R} {R'}{H \over H'} \\
&=& {2 \over 3(1+f_R)} {1 \over 4\Epar'+\Epar''}{ \Epar \over \Epar'} \left( y''- y' { 4\Epar'' + \Epar''' \over 4 \Epar' +\Epar''} \right)\,, \nonumber
%\,,\\
% f_R &=& {y' \over 3(4 \Epar' + \Epar'')}
\end{eqnarray}
is most closely linked with the phenomenology.  $\Bpar$ is a strongly
growing function in our solutions and in the high curvature limit
has a growth rate
\begin{equation}
p_\Bpar \equiv {\Bpar' \over \Bpar}
\label{eq:Bgrowth}
\end{equation}
given by
$p_\Bpar=3+p_+$, if the growing mode dominates,
and $p_\Bpar=-3w$, if the particular mode dominates.

We will therefore characterize
solutions with a given expansion history family by  $ \Bpar_0 \equiv \Bpar(\ln a=0)$.
If $\Bpar_0=0$ and the background
expansion is given by $w=-1$ then $f(R)=$ const., $\Bpar(\ln a)=0$ and
the model has a true cosmological constant.
More generally $\Bpar=0$ will correspond in linear theory
to the dynamics of a dark energy component for scales above the dark energy sound horizon.

\begin{figure}[htbp]
  \begin{center}
  \epsfxsize=3.3truein
    \epsffile{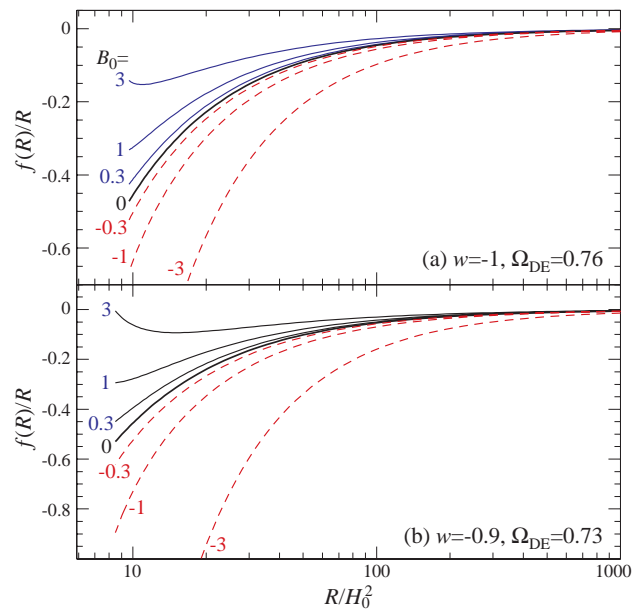}
    \caption{\footnotesize Every expansion history that can be parameterized by a dark
    energy model with $\rho_{\rm DE}(\ln a)$ can be reproduced by a one
    parameter family of $f(R)$ models, indexed by $\Bpar_0 \propto f_{RR}/(1+f_R)$ at the
    present epoch (left end point of curves),
    that approaches the Einstein-Hilbert action in the high curvature
    limit.   (a) $\Lambda$CDM expansion history
    ($w=-1$, $\Omega_{\rm DE}=0.76$, $h=0.73$). (b) Dynamical dark energy
    expansion history
   ($w=-0.9$, $\Omega_{\rm DE}=0.73$, $h=0.69$).
}
\label{fig:background}
\end{center}
\end{figure}

In Fig.~\ref{fig:background}, we plot the family of $f(R)$ models that match
 two representative
expansion histories parameterized by $(w, \Omega_{\rm DE},h)$.   These models are
chosen to be consistent with current WMAP CMB data \cite{Speetal06}
 and span a range that is consistent
with supernovae acceleration measures.
   Given the similarity between these models, we will
take the $w=-1$ $\Lambda$CDM expansion history for illustrative purposes below.

The linear perturbation analysis that follows does not require the matching to the specific
expansion histories parameterized by $(w, \Omega_{\rm DE}, h)$ here.
This reverse engineering is only a device to find
observationally acceptable $f(R)$ models.  What is required is that the background solutions
provide $H(\ln a)$ and $\Bpar(\ln a)$.   On the other hand,
linear perturbation theory does inform the choice of a background solution.  We shall
find in \S \ref{sec:stability} that the $\Bpar<0$ branch of the family is unstable to linear perturbations in the high curvature regime.

\section{Linear Perturbation Equations}
\label{sec:linear}

The modified Einstein equations (\ref{eq:metricvar}) represent a fourth-order set of
differential equations for the two metric perturbations $\Psi$ and $\Phi$ in
the presence of matter density and momentum fluctuations $\delta$ and $q$.
To solve this system of equations, we introduce auxiliary parameters to
recast it as a larger set of second-order differential equations.  It is numerically and pedagogically advantageous to choose
these auxiliary parameters so that their effect vanishes at large scales
and early times.

On superhorizon scales ($k/aH\ll 1$), the evolution of metric perturbations must be consistent with
the background evolution provided that the background solution is valid, i.e.~that
fluctuations about it are stable.
Bertschinger \cite{Ber06} showed that the familiar
conservation of the curvature fluctuation on comoving hypersurfaces ($\zeta'=0$)
for adiabatic fluctuations in a flat universe
applies to any metric based modified gravity model that obeys energy momentum conservation
$\nabla^\mu T_{\mu \nu}=0$.
The gauge transformation into the Newtonian gauge
\begin{equation}
\zeta = \Phi + H q
\label{eq:gaugetransform}
\end{equation}
implies
\begin{equation}
\zeta' = \Phi' + H' q + H q'  = 0\,,  \,\,\, (k=0) \,,
\end{equation}
and momentum conservation requires
\begin{equation}
H q' = -\Psi
\label{eq:momentumconservation}
\end{equation}
so that
\begin{equation}
\Phi' - \Psi + H'q = 0\,,  \,\,\, (k=0) \,.
\label{eq:0izerok}
\end{equation}
Combining Eq.~(\ref{eq:momentumconservation}) and (\ref{eq:0izerok}) yields
a second order differential equation for the Newtonian metric perturbations
\cite{HuEis99}
\begin{eqnarray}\label{eq:Bert}
\Phi'' - \Psi'  -{H'' \over H'}\Phi' - \left( {H' \over H} -{H'' \over H'} \right) \Psi = 0
\,,  \,\,\, (k=0) \,.
\end{eqnarray}
The evolution of the metric fluctuations must in this way be consistent with the expansion history
defined by $H$.
Note that this equation applies to any modified gravitational scenario that satisfies
the required conditions.  The DGP braneworld acceleration model \cite{DvaGabPor00}
represents another valid application \cite{SawSonHu06}.
What does require a specification of a theory is
the relation between $\Phi$ and $\Psi$.  Under general relativity and assuming
$T_{\mu\nu}$ takes the matter-only form of Eq.~(\ref{eq:Tmunu}), the closure
relation is $\Phi = -\Psi$ and Eq.~(\ref{eq:Bert}) in fact applies on all scales.  With a dynamical
dark energy component in $T_{\mu\nu}$, it applies above the dark energy sound horizon
\cite{Hu98}.

To capture the metric evolution of $f(R)$ models for $k \ne 0$, let us introduce two
 parameters:
 $\thepar$ the deviation from $\zeta$ conservation, Eq.~(\ref{eq:0izerok})
 \begin{eqnarray}\zeta' = \Phi' -\Psi + H'q &=& { H' \over H} \left( {k \over a H} \right)^2 \Bpar \thepar \,,
 \label{eq:thepardef}
\end{eqnarray}
 and $\epspar$ the deviation from the superhorizon metric evolution Eq.~(\ref{eq:Bert})
\begin{eqnarray}
&&\Phi'' - \Psi'  -{H'' \over H'}\Phi' - \left( {H' \over H} -{H'' \over H'} \right) \Psi  \nonumber\\
% typo in v1 && \qquad =  \left( {k \over a H} \right)^2{\Epar' \over 4\Epar' + \Epar''} \Bpar^2 \epspar \,.
&& \qquad =  \left( {k \over a H} \right)^2 \Bpar \epspar \,.
\label{eq:epspardef}
\end{eqnarray}
The coefficients in front of the deviation parameters are chosen to bring out the
fact that  their effect vanishes as $k/aH \rightarrow 0$ and $\Bpar \rightarrow 0$ so long as
the dynamics guarantees stable behavior of the parameters themselves (see \S \ref{sec:stability}).

 The Einstein equations can then be recast as a second order differential equation
 for $\epspar$ and constraint equations for the other metric variables.
 Since $\epspar$ itself contains second derivatives of the fundamental
 metric perturbations $\Phi$ and $\Psi$, the equations are implicitly fourth order.
It will be convenient to separate out two linear combinations of the underlying metric
fluctuation
\begin{equation}
\Phi_- = {1\over 2}(\Phi - \Psi) \,,  \quad S = 2\Phi+\Psi \,,
\end{equation}
and a reduced mass scale or rescaling of $G$
\begin{equation}\label{eq:mu2}
\tilde \mu^2(\ln a) = {\mu^2 \over 1+f_R(\ln a)} \,.
\end{equation}
In terms of these variables, the $0i$ component of the Einstein equations becomes
a dynamical equation for $\epsilon$
 \begin{eqnarray}
 \label{eq:epspardynamics}
\Bpar( \epspar' + G_1 \epspar) &=& {1\over 3} G_2 S   -{1\over 3} (S - 2\Phi_-)
+  {\Bpar \over 6}{\Epar' \over \Epar}(S -2\Phi_-)
\nonumber  \\
&& +\left(  {1\over 2}{\Epar' \over \Epar}  + {1\over 3} {\tilde \mu^2 \rho \over H^2} \right)  H q
\\
&&
+ \left[ {\tilde \mu^2 \rho \over H^2}\left( 4 + {\Epar'' \over \Epar'} \right)
- {1\over 2}{\Epar' \over \Epar} \left( {k \over a H} \right)^{2} \right]  \thepar  \,, \nonumber
\end{eqnarray}
where
\begin{eqnarray}
G_1 &=& 1 - {\Epar' \over \Epar} + 2 {\Epar'' \over \Epar'} - {4 \Epar'' + \Epar''' \over 4 \Epar' + \Epar''} +
{1 \over 2}{\Epar' \over \Epar} \Bpar + 2 {\Bpar' \over \Bpar}\,, \nonumber\\
G_2 &=& 4 + {\Epar'' \over \Epar'} + {\Bpar' \over \Bpar} - G_1\,.
\end{eqnarray}
The metric fluctuations $S$ and $\Phi_-$ act as sources to $\epspar$ which then
feed back into their evolution weighted by $(\Bpar k/aH)^2$.
To complete this system, the dynamics of $\thepar$ are supplied by the
derivative of its definition Eq.~(\ref{eq:thepardef}) combined with Eq.~(\ref{eq:epspardef})
\begin{eqnarray}
\thepar' + \left( -2 -{3\over 2}{\Epar' \over \Epar} + {\Bpar' \over \Bpar}\right) \thepar& = &2{ \Epar \over \Epar'} \epspar\,.
\label{eq:thepardynamics}
\end{eqnarray}
Eqs. (\ref{eq:epspardynamics}) and (\ref{eq:thepardynamics}) can also be combined to eliminate
$\thepar$ leaving a second-order differential equation for $\epspar$.  This combined relation
may alternately be derived directly from the trace of the $ij$ component of the Einstein equations.

As in general relativity, the remaining Einstein equations become constraint equations given
the dynamical variables $\Delta, Hq, \epspar, \thepar$.
The $00$ equation may be expressed as the modified Poisson equation,
\begin{eqnarray}
  2\Phi_- -{\Bpar \over 2}{\Epar' \over \Epar}{\Epar' \over 4\Epar' + \Epar''}
(S + 3 \Bpar \epspar) %\nonumber\\
%\qquad &&
= {\tilde \mu^2 a^2 \rho \over k^{2}} \Delta\,,
\end{eqnarray}
where $\Delta$ is the density perturbation in the comoving gauge
\begin{equation}
\Delta = \delta - 3 H q \,,
\end{equation}
and the trace-free $ij$ component becomes
\begin{eqnarray}
\Phi + \Psi &=&
{2 \over 3}( S - \Phi_-)
\nonumber\\
&=& {\Bpar \over 2}{\Epar' \over \Epar} (H q)
- \Bpar \left( { k \over a H} \right)^2 \\ &&
\times \Big[ {1\over3}{ \Epar' \over 4\Epar' + \Epar''} ( S  + 3 \Bpar \epspar) + {1 \over 2}{\Epar' \over \Epar} \Bpar \thepar \Big] \,.
\label{eq:anisotropy}
\nonumber
\end{eqnarray}
Note that as $\Bpar \rightarrow 0$ and $\tilde \mu \rightarrow \mu$ these constraint
equations become the usual Poisson and anisotropy
equations.  In particular for $\Bpar = 0$, the dark energy-closure relation $\Phi=-\Psi$ is recovered.

Finally, the conservation laws provide the dynamics for the matter fluctuations
and are unmodified by $f(R)$
\begin{eqnarray}
\Delta'=
\left( \frac{k}{aH} \right)^2 Hq-3\zeta'\,,
\end{eqnarray}
\begin{eqnarray}
H q'=-\Psi = -\frac{1}{3}(S-4\Phi_-).
\label{eq:momentumconservation2}
\end{eqnarray}
The impact of the modification to gravity comes from the metric evolution.  Eq.~(\ref{eq:thepardef})
implies
\begin{eqnarray}
\zeta' = {H' \over H} \left( {k \over a H} \right)^2 \Bpar \thepar \,.
\label{eq:zetaprime}
\end{eqnarray}
In fact directly integrating Eq.~(\ref{eq:zetaprime}) and checking for consistency
between $H q$ defined through Eq.~(\ref{eq:gaugetransform}) and
Eq.~(\ref{eq:momentumconservation2}) tests  the numerical accuracy
of solutions.

Along with initial conditions for each of the fluctuations, these equations provide a complete
and exact description of scalar linear perturbation theory in $f(R)$ gravity for a matter-only universe.

\section{Stability at High Curvature}
\label{sec:stability}

The fourth-order nature of the linear perturbation equations derived in the previous
section raises the question of stability in the high-curvature limit to general relativity
\cite{DolKaw03,Zha05,Woo06,SieWal06}.
 Strongly unstable metric fluctuations can create order unity effects
 that invalidate the background expansion history.

The key equations for stability
 are (\ref{eq:epspardynamics}) and (\ref{eq:thepardynamics}) which describe
the evolution of the deviation parameters.  Consider the high redshift  limit of high curvature where
$|\Bpar| \rightarrow 0$ and wavelengths of interest are well outside the horizon $k/aH \ll 1$.  In this
limit the evolution equations simplify to
\begin{equation}
\epspar'' + \left( {7 \over 2} + 4p_\Bpar \right) \epspar' + {2 \over \Bpar} \epspar = {1 \over \Bpar} F(\Phi_-,S,Hq) \,,
\label{eq:stabilityeqn}
\end{equation}
where $F(\Phi_-,S,Hq)$ is the source function for the deviation $\epsilon$ and recall
$p_\Bpar$ is the growth index of $B$ from Eq.~(\ref{eq:Bgrowth}).
  The details of $F$ are not important for the stability analysis
  other than that it provides a source that is of order
the perturbation parameters that are its arguments.  Under the assumption that
the general relativistic solution is stably recovered in this limit, it acts as an external source
to the deviations.  The stability question can be phrased
as whether $\epspar$ remains self-consistently of order these sources or grows and prevents
the recovery of the solutions.

The stability equation (\ref{eq:stabilityeqn}) has the peculiar feature that the frequency squared
$2/B$
diverges as $|B| \rightarrow 0$ independently of $k$, resembling a divergent real or imaginary
mass term.
Evolution of $\epsilon$ can occur on
a time scale much shorter than the expansion time.  If $B < 0$, $\epspar$ is highly
unstable and deviations will grow exponentially.  If $B>0$, $\epspar$ is highly stable and
is driven to the value required by the source function $\epsilon = F/2$.  This short time scale
behavior can also be seen directly in the 4th order form of the Einstein equation. The
trace of the $ij$ equation or the derivative of the $0i$ equation have their 4th order
terms multiplied by the small parameter $B$.

Thus despite the apparent recovery of general relativity in the action at high curvature $R$,
the general-relativistic solutions to linear perturbation theory are not recovered for $B<0$.
In terms of $f(R)$, $B \propto f_{RR}$ in this limit
and hence models like \cite{Caretal03}
\begin{equation}
f(R) = - {M^{2+2n} \over R^{n}}\,, \quad (n>0)\,,
\end{equation}
and \cite{Zha05}
\begin{equation}
f(R) = -M^2 \exp(-R/\lambda M^2)\,,
\end{equation}
are included in this class of unstable models. 

The instability causes any finite patch of a universe that starts at high curvature
 to break away from the background
solution into either a low curvature solution or a singularity.   The low curvature
$R \ll G\rho$ solutions on the other hand are stable and in fact correspond to the
background expansion histories studied by \cite{AmePolTsu06a,Beaetal06}.  
However these expansion histories have gravity modified throughout the matter
dominated epoch and in particular $a \propto t^{1/2}$.  They 
 produce phenomenology at high redshift that
would violate constraints from the CMB.
We will omit them from further consideration below.

\section{Metric Evolution Solutions}
\label{sec:evolution}

In this section we discuss the numerical solutions of the linear perturbation equations on
the stable $B>0$ branch.   To expose the underlying features of the solutions,
we examine the relevant limiting cases below.  We begin with the initial epoch where
$|B| \ll 1$ and the fluctuations are superhorizon sized $k/aH \ll 1$.  We then
examine large scale or ``superhorizon" modes where $B^{1/2} k/aH \ll 1$ whose evolution is
completely determined by the background expansion history and the form of $f(R)$. 
Finally we track the evolution of small scale or ``subhorizon" modes until $B^{1/2} k/aH \gg 1$
where their evolution reaches the simple form implied by quasistatic equilibrium.

\subsection{Initial Conditions}

On the stable $B>0$ branch, we can set the initial conditions when $B \ll 1$ and
the mode is superhorizon sized, $k/aH \ll 1$.  In this case, the initial conditions for
the normal fluctuation parameters follow the general-relativistic expectation
\begin{eqnarray}
\Phi_i &=& {3 \over 5} \zeta_i \,, \nonumber\\
\Psi_i  &=& -\Phi_i \,, \nonumber\\
\Delta_i &=& {2 \over 3} \left( {k \over a H} \right)^2 \Phi_i \,,\nonumber\\
H q_i &=& {2 \over 3} \Phi_i \,,
\end{eqnarray}
where $\zeta_i$=const. is the initial comoving curvature.
These relations also imply $\Phi_- = \Phi$ and $S = \Phi$ with vanishing
first derivatives initially.

Detailed balance gives the deviation parameters as
\begin{eqnarray}
\thepar_i &=& {1 \over 9}p_B \Phi_i \,,\nonumber \\
\epspar_i &=& -{3 \over 2} \left( {5 \over 2} + p_B  \right) \thepar_i \,,
\label{eq:initialdeviation}
\end{eqnarray}
and the high frequency term in their evolution equations ensures that they stay locked
to these relations until $B$ becomes non-negligible.

\subsection{Superhorizon Evolution}

Given that $\thepar$ and $\epspar$ are locked to the initial values
of Eq.~(\ref{eq:initialdeviation}) when $B\ll 1$,
their definitions in Eq.~(\ref{eq:thepardef}), (\ref{eq:epspardef})
imply that they have negligible effect on the evolution of the metric fluctuations $\Phi$, $\Psi$.
This remains true even as the mode evolves into the $B \sim 1$ regime 
if $k/aH \ll 1$.
In particular, the anisotropy relation of Eq.~(\ref{eq:anisotropy})
 becomes
 \begin{equation}
\Phi + \Psi = B H' q
\label{eq:phipsiclosure}
\end{equation}
and closes the general relation for superhorizon metric fluctuations Eq.~(\ref{eq:Bert}):
\begin{eqnarray}
&&\Phi'' + \left( 1 - {H'' \over H'} + {B' \over 1-B} + B {H' \over H} \right) \Phi'\\
&&\qquad  + \left( {H' \over H} - {H'' \over H'} + {B' \over 1-B} \right) \Phi =0 \,,\,\,\, (k=0)\,.\nonumber
\label{eq:Bertclosed}
\end{eqnarray}
The evolution of $\Phi$
is completely determined by the background evolution and the specification of $f(R)$.
Formally, this solution also applies to the unstable branch $B<0$ at $k=0$ but is only
valid at finite $k$ for large $|B|$.  The point at which $B=1$ is a regular singular point for typical
$B(\ln a)$ and so $\Phi$
evolves smoothly through it.  $\Phi$ will grow on the expansion time scale if
\begin{equation}
{H' \over H} - {H'' \over H'} + {B' \over 1-B} < 0 \,.
\end{equation}
Growth is typically a transient phenomenon at the onset of acceleration given that the presence
of matter makes $H'/H - H''/H'$ positive.
For example, if the expansion rate approaches the de Sitter case of a constant
in the future  $H'/H \rightarrow 0$ and the solution to Eq.~(\ref{eq:Bertclosed}) becomes
\begin{equation}
\Phi = {C_1 \over a} + {C_2 \over a} \int da {(1-B) H'} \,, \,\,\, (H'/H \rightarrow 0)\,,
\label{eq:deSitter}
\end{equation}
where $C_1$ and $C_2$ are constants.
This implies decaying solutions unless $B H'$ grows.   $H'$ decays and, since in de Sitter
 $R\rightarrow$ const,  $B$ should asymptote to a constant.  Eq.~(\ref{eq:deSitter}) then
 implies  $\Phi \propto 1/a$.
 Note that this stability analysis differs from treatments that take a pure de Sitter expansion
 with no matter since that assumption forces a closure relation of $\Phi=\Psi$
 (c.f. \cite{Far05a,Far05b}).

\begin{figure}[htbp]
  \begin{center}
  \epsfxsize=3.3truein
        \epsffile{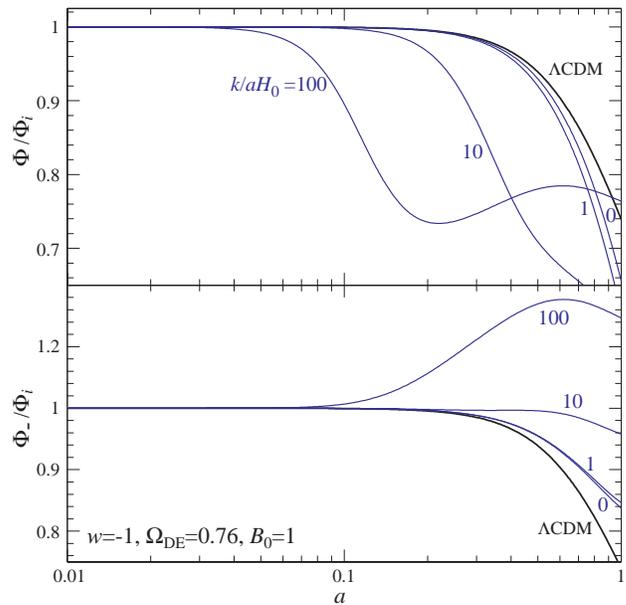}
    \caption{\footnotesize Evolution of 
    metric fluctuations $\Phi$ (upper panel) and $\Phi_{-}$ (lower panel) for $B_{0}=1$
    and a $\Lambda$CDM expansion history.  The different
    closure relations on super and sub-horizon scales for $\Psi$, Eqs.~(\ref{eq:phipsiclosure})
    and (\ref{eq:Sclosure}), lead to qualitatively different evolution for the two limits
    with a transition region in between.
    $\Phi_{-}$, which controls effects in the CMB and enters directly into the Poisson equation,
    has a scale-dependent growth that makes it increasingly larger
    than the $\Lambda$CDM prediction at high $k$.  Results for other
    values of $B_{0}$ can be scaled from this figure by noting that the transition occurs
    when $k/aH \approx B^{-1/2}$.
}
\label{fig:time}
\end{center}
\end{figure}

An example of the superhorizon evolution of the metric is shown in Fig.~\ref{fig:time}
($k=0$ curves) for a model with $B_{0}=1$.   While $\Phi$ is monotonically smaller
than the $B=0$ $\Lambda$CDM prediction, $\Phi_{-}$ is monotonically larger due
to the closure relation between $\Phi$ and $\Psi$ of Eq.~(\ref{eq:phipsiclosure}).

\subsection{Subhorizon Evolution}

For subhorizon scales where $k/aH \gg 1$,
Eqs.~(\ref{eq:epspardynamics}) and (\ref{eq:thepardynamics}) form
an oscillator equation whose frequency scales as $k/aH$.  Therefore
 the amplitude of $\epspar$ is driven to zero when compared with $\Phi_-$.  When combined with the Poisson and anisotropy  equation, this requires
\cite{Zha05}
\begin{equation}
\lim_{B k/aH \rightarrow \infty} S \rightarrow0 \,, \quad \Psi \rightarrow -2 \Phi \,, \quad
\Phi_- \rightarrow {3 \over 2}\Phi
\label{eq:Sclosure}
\end{equation}
The Poisson equation then takes the simple form
\begin{equation}
k^2 \Phi_- = {1 \over 2}{\tilde \mu^2 a^2 \rho} \Delta
\end{equation}
and the conservation laws become
\begin{eqnarray}
\Delta' &=& \left( { k \over a H} \right)^2 Hq\,, \nonumber\\
Hq' & = & {4 \over 3} \Phi_- \,,
\end{eqnarray}
where we have dropped temporal derivatives when compared with spatial gradients where
appropriate.
This system describes a scale free evolution for $\Phi_-$ or $\Delta$.
The transition between these two scale-free regimes occurs when
$(k/aH) \sim B^{-1/2}$.  This scale- and time-dependent transition leads
to a scale-dependent growth rate.  Unlike for $\Phi$, $\Phi_{-}$ has monotonically
enhanced power as $k$ increases on the  $B>0$ branch.
Because of the time dependence of the transition, the total growth to $z=0$
continues to increase with $k$ even for $k/aH \gg B^{-1/2}$.

In Fig.~\ref{fig:time}, we show the full numerical solution from the initial conditions through
the super- to the sub-horizon evolution for a few representative modes.

\section{Power-Spectra Observables}
\label{sec:observables}

The scale dependences of the linear growth rate of metric and density perturbations change
predictions for cosmological power spectra in the linear regime.
Let us make the usual assumption that the initial spectrum of fluctuations in the comoving curvature
is given by a power law.  For a starting epoch during matter domination, this power
law is modified by the usual matter-radiation transfer function $T(k)$
\begin{equation}
{k^3 P_{\zeta_{i}} \over 2\pi^2} = \delta_\zeta^2 \left( {k \over k_{\rm n}} \right)^{n-1}
T^{2}(k) \,,
\end{equation}
where $\delta_\zeta$ is the rms amplitude at the normalization scale
$k_{\rm n}$.

The modifications to the CMB depend on the scale-dependent change in the
potential growth rate
\begin{equation}
G(a,k) = {\Phi_{-}(a,k) \over \Phi_{-}(a_{i},k)} \,,
\label{eq:potentialgrowth}
\end{equation}
through the Integrated Sachs-Wolfe (ISW) effect.
This effect comes from the differential redshift that CMB photons suffer
as they transit
the evolving potential.  It contributes to the angular power spectrum of temperature
anisotropies as
\ba
C^{\rm II}_{l}=4\pi \int \frac{{d}k}{k}
[I^{\rm I}_{l}(k)]^{2} {9 \over 25} {k^{3}P_{\zeta_{i}} \over 2\pi^{2}}\,,
\label{eqn:ISWCl}
\ea
where
\ba
I^{\rm I}_{l}(k)=2 \int dz { dG \over dz}
j_{l}(kD)\,.
\ea
Here $D=\int dz/H$ is the comoving distance out to redshift $z$.

\begin{figure}[htbp]
  \begin{center}
  \epsfxsize=3.3truein
    \epsffile{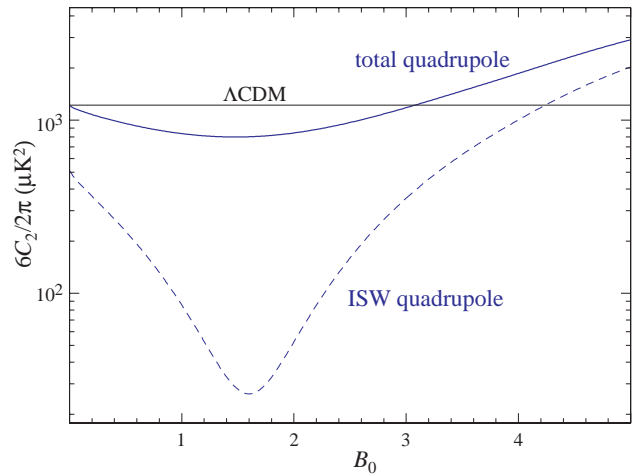}
    \caption{\footnotesize  CMB quadrupole power $6 C_2/2\pi$ contributed by the modified
    ISW effect (dashed curve) and total (solid curve) as a function of $\Bpar_0$ in the $\Lambda$CDM
    expansion history.  For reference, the $\Lambda$CDM total quadrupole is also shown
    (horizontal line). The change in the growth of the potential causes a near nulling of
    the ISW effect at $\Bpar_0 \approx 3/2$ and a substantial reduction of power between
    $0.2 \la \Bpar_0 \la 2.5$.
}
\label{fig:quad}
\end{center}
\end{figure}

In Fig.~\ref{fig:quad}, we show the quadrupole
power as a function of $\Bpar_0$ contributed by the ISW effect as well as the
total quadrupole.  Power in the quadrupole arises near scales of $k/H_{0} \sim 10$
and so the weak evolution of $\Phi_{-}$ shown in Fig.~\ref{fig:time} reduces
it near $B_{0}\sim 1$.
In fact there is a minimum around $\Bpar_0 \approx 3/2$, where
the ISW effect is a negligible contributor to the power, and
a substantial reduction for $0.2 \la \Bpar_0 \la 2.5$.   Further reduction of large-scale
power can be achieved by changing the initial power spectrum to simultaneously
suppress horizon scale power in the Sachs-Wolfe
effect from recombination.   Hence these models provide the opportunity
to bring the predicted ensemble-averaged quadrupole closer to the measurements
on our sky \cite{Speetal06}.
Models with $\Bpar_0 \ga 3$ produce an excess of large-angle anisotropy and exacerbate
the tension with the data.
Note however that due to sample variance, changes in the likelihood will be
small.
We will address constraints on the models in a separate work.

\begin{figure}[htbp]
  \begin{center}
  \epsfxsize=3.3truein
    \epsffile{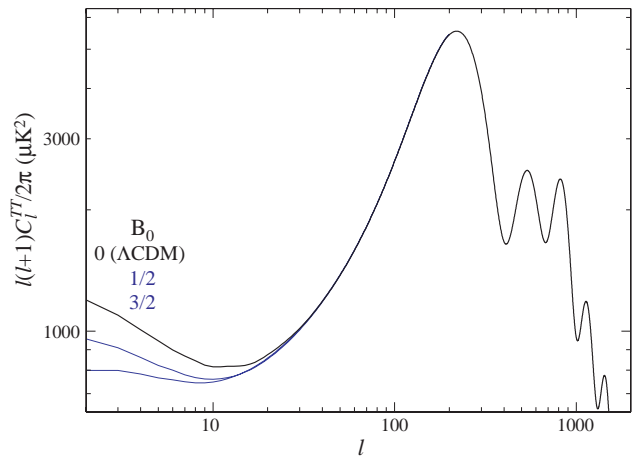}
    \caption{\footnotesize CMB angular power spectra for the $\Lambda$CDM expansion history
    for $\Bpar_0=0$ ($\Lambda$CDM), 1/2, 3/2. Power in the low multipoles is lowered by the
    reduction in the ISW effect. Power  at the high multipoles of the acoustic peaks is left
    unchanged.
}
\label{fig:cl}
\end{center}
\end{figure}

We show the full spectrum of
temperature anisotropy $C_{l}^{TT}$ in Fig.~\ref{fig:cl} for a few representative values of $B_{0}$.
Given that the changes to the power spectrum occur mainly at the lowest multipoles,
WMAP constraints on the amplitude of the peaks can be directly translated into
a normalization of the power spectrum on scales corresponding to the acoustic peaks.
For the $\Lambda$CDM
expansion history of $w=-1$, $\Omega_{\rm DE}=1-\Omega_m=0.76$,
$h=0.73$ the normalization from WMAP is $\delta_{\zeta}=4.52 \times 10^{-5}$ for
an optical depth to reionization of $\tau=0.092$.  We further take a tilt of $n=0.958$, $\Omega_b h^2=0.0223$.

The WMAP normalization then allows us to predict the matter power spectrum today.
Let us define the density growth rate
\begin{equation}
D_G(a,k) = {\Delta(a,k) \over \Delta(a_{i},k) }a_{i}
\label{eq:densitygrowth}
\end{equation}
such that $D_G=a$ before $f(R)$ effects become important.
In $f(R)$ models the potential and density growth rates Eq.~(\ref{eq:potentialgrowth})
and (\ref{eq:densitygrowth}) can differ non-trivially due to the time dependent
 $(1+f_R)$ rescaling of $G$ in
the Poisson equation.
The linear power spectrum then becomes
\begin{eqnarray}
%\Delta^{2}_{m}(k,a) &=&
{k^{3} \over 2\pi^{2}} P_{L}(k,a)
&=& {4 \over 25} D_G^{2}(a,k)
{k^{4} \over \Omega_{m}^{2}H_{0}^{4}}  {k^{3}P_{\zeta_{i}} \over 2\pi^{2}}  \,.
\end{eqnarray}

\begin{figure}[htbp]
  \begin{center}
  \epsfxsize=3.3truein
    \epsffile{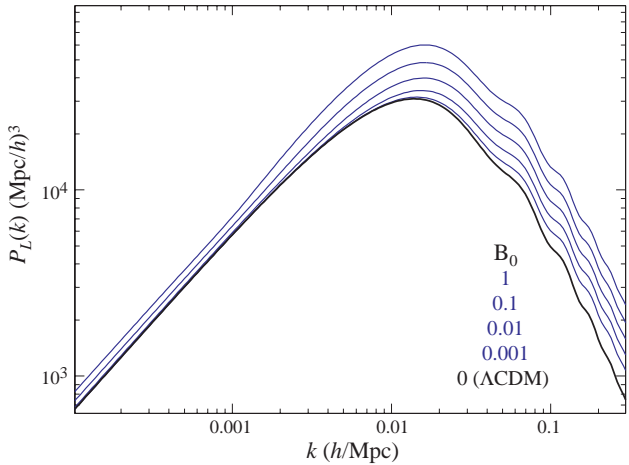}
    \caption{Linear matter power spectrum for several values of $\Bpar_0$ in the $\Lambda$CDM
    expansion history.   The change in the amplitude of the power at high
    $\Bpar_0 \ga 0.1$ is nearly degenerate with galaxy bias.
    Smaller values of $0.001 \la \Bpar_0 \la 0.1$ change
    the shape of the linear power spectrum  at a potentially observable level.  All spectra
    are normalized to the WMAP anisotropy from recombination.
}
\label{fig:power}
\end{center}
\end{figure}

We show $P_L(k)$ for several choices of $\Bpar_0$ in Fig.~\ref{fig:power}.   Despite the
large change in amplitude at  high $k$, the high $\Bpar_0$ models cannot be automatically
ruled out by galaxy clustering data
since the nearly multiplicative shift can be mimicked by galaxy bias.
Likewise, non-linear measurements of the mass power spectrum through the cluster
abundance, Lyman-$\alpha$ forest, and cosmic shear  also cannot be straightforwardly
applied.
 As the local curvature exceeds the background curvature in
collapsed dark matter halos one would expect the gravitational dynamics to return
to Newtonian.   For this reason, our predictions are restricted to the linear regime
at $k \la 0.1 h$ Mpc$^{-1}$.
We intend to explore these issues further in a future work.

Finally the cross correlation between the ISW effect and the angular power spectra
of galaxies is markedly different in these models and potentially excludes large
$\Bpar_0$ solutions.
The angular power spectra of galaxies is given in the linear regime by
\ba
C^{g_{j}g_{j}}_{l}=4\pi \int \frac{dk}{k}
[I^{g_{j}}_{l}(k)]^2 {4 \over 25} {k^{4} \over \Omega_{m}^2H_{0}^{4}}
 {k^{3}P_{\zeta_{i}} \over 2\pi^{2}},
\ea
where
\ba
I^{g_{j}}_{l}(k)=\int {d}z D_{G}(a,k) {n_{j}(z)b_{j}(z)}
j_{l}(kD)\,,
\ea
$n_{j}(z)$ is the galaxy redshift distribution normalized to
 $\int dz n_{j} = 1$, and $b_{j}(z)$ is the galaxy bias.

The cross correlation between the CMB ISW effect and galaxies becomes
\ba
C^{g_{j}{\rm I}}_{l}=4\pi \int \frac{dk}{k}
I^{g_{j}}_{l}(k)I^{\rm I}_{l}(k) {6 \over 25} {k^{2} \over \Omega_{m}H_{0}^{2}}
 {k^{3}P_{\zeta_{i}} \over 2\pi^{2}} \,.
\ea
The correlation coefficient between the total temperature anisotropy and the
galaxies is given by
\ba
R_l \equiv { C_l^{g_j {\rm I}} \over \sqrt{C_l^{TT} C_l^{g_j g_j}}}
\ea
and is independent of the galaxy bias if it is slowly varying with redshift.
We have neglected galaxy magnification bias which leads to an additional
source of correlation.

For definiteness,
we assume that the galaxy sets come from a net galaxy distribution of
\ba
n_g(z)\propto z^2 e^{-(z/1.5)^2}\,,
\ea
which is further partitioned by photometric redshift into several galaxy samples,
\ba
n_j(z) \propto n_g(z)
\left[ {\rm erfc}\left(\frac{z_{j-1}-z}{\sqrt{2}\sigma(z)}\right)
-{\rm erfc}\left(\frac{z_{j}-z}{\sqrt{2}\sigma(z)}\right)\right]\,,
\label{eq:photozpartition}
\ea
where erfc is the complementary error function
and $\sigma(z)=0.03(1+z)$ reflects the effect
of photometric redshift scatter.

\begin{figure}[htbp]
  \begin{center}
  \epsfxsize=3.3truein
    \epsffile{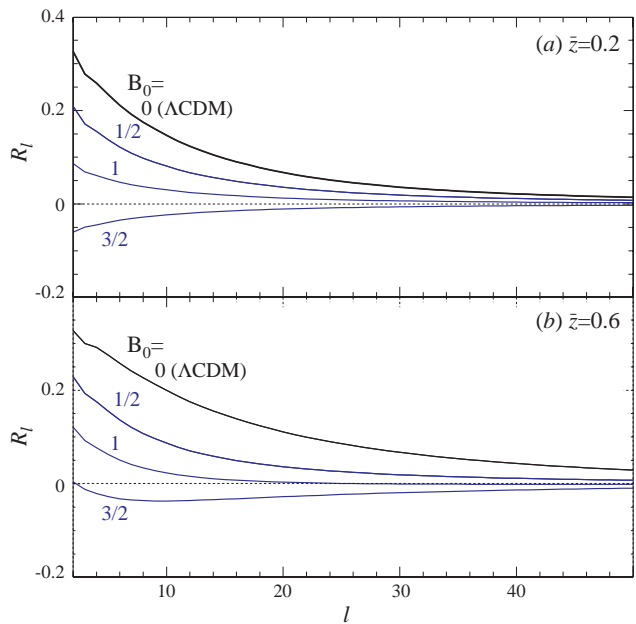}
    \caption{Cross correlation coefficient between the CMB and galaxies in the $\Lambda$CDM
    expansion history.  Shown are two
    representative redshift bins centered around $\bar z=0.2$ and $0.6$ with
    $\Bpar_0=0$ \,($\Lambda$CDM), 1/2, 1, 3/2.
    The cross correlation is substantially reduced for $1/2 \la \Bpar_0 \la 1$ and becomes
    negative for $\Bpar_0 \ga 3/2$.
}
\label{fig:galisw}
\end{center}
\end{figure}

Fig.~\ref{fig:galisw} shows the correlation coefficient for several values of $\Bpar_0$.
We take two redshift bins from Eq.~(\ref{eq:photozpartition}) partitioned by
$z_j=0,0.4,0.8$ to achieve effective redshifts of $\bar z = 0.2$ and $0.6$.
Current observations constrain the correlation near $l \sim 20$ corresponding
to scales which are an order of magnitude smaller than those contributing to the
ISW quadrupole.
Between $1/2 \la \Bpar_0 \la 1$ the galaxy-ISW correlation is substantially reduced.
For $\Bpar_0 \ga 3/2$, galaxies are in fact anti-correlated with the CMB since
$\Phi_-$ grows on the relevant scales.   A loose bound from the observed
correlation would therefore be $\Bpar_0 \la 3/2$ at the 
significance levels of the reported detections (e.g. \cite{BouCri04,FosGaz04,Scretal04,Noletal03,AfsLohStr04,Giannantonio06})
 but we expect more detailed modeling
to yield better constraints in the future.   It is likely that
a significant reduction of the large angle anisotropy from this mechanism could
be excluded unless other sources, such as magnification bias,
 can generate the observed positive correlation.

\section{Discussion}
\label{sec:discussion}

We have studied the evolution of 
linear cosmological perturbations in
$f(R)$ models for
modified gravity in the physical---or Jordan---frame.  Here the gravitational dynamics
are fourth order and the matter is minimally coupled and separately conserved.
For models that recover the Einstein-Hilbert action at high curvature $R$,
we find that for each expansion history specified by $H(\ln a)$ there exists two
branches of $f(R)$ solutions that are parameterized by $B \propto f_{RR}$, the
second derivative of $f(R)$.   For $B<0$, which includes most models
previously considered \cite{Caretal03,Zha05}, there is a short-timescale
instability that prevents recovery of the general-relativistic expectations at high
curvature that is important for maintaining agreement with CMB measurements.

For the stable $B>0$ branch, $f(R)$ models predict a rich set of linear phenomena
that can be used to test such deviations from general relativity.  For example,
large $B \sim 1$ models lower the large-angle anisotropy of the CMB and may be
useful for explaining the low quadrupole observed on our sky.  They also predict
qualitatively different correlations between the CMB and galaxy surveys which
may provide the best upper limit on the deviations currently available.
Smaller deviations in $B$ are observable at smaller scales through changes
to the shape of the linear power spectrum.   In the limit that $B \rightarrow 0$ and
the expansion history is given by $\Lambda$CDM, linear perturbations in
$f(R)$ models approach
the general relativistic predictions exactly.   We intend to examine constraints
on $f(R)$ models in a future work.

More generally, this class of phenomenological $f(R)$ models provides insight
on the types of deviations that might be expected from alternate metric theories of gravity
 in the linear regime.  Conservation of the matter stress-energy tensor severely
 restricts the form of allowed deviations on both super- and sub- horizon scales
 \cite{Ber06,SawSonHu06}.    Even if these $f(R)$ models prove not to be
 viable as a complete alternate theory of gravity that includes solar-system tests,
 they may serve as the basis for a ``parameterized post-Friedmann" description
 of linear
 phenomena that parallels the parameterized post-Newtonian description of small-scale tests.

\vspace{1.cm}

\noindent {\it Acknowledgments}: We thank Sean Carroll, Mark Hindmarsh,
Michael Seifert, Kendrick Smith, Bob Wald, and the participants
of the Benasque Cosmology Workshop and the Les Houches Summer School for useful conversations.
This work was supported by the
U.S.~Dept. of Energy contract DE-FG02-90ER-40560. IS and WH are
additionally supported by the David and Lucile Packard Foundation.
This work was carried out at the KICP under NSF PHY-0114422.
\hfill
\bibliography{fofr}

\end{document}